\pdfoutput=1
\documentclass[10pt, conference, compsocconf]{IEEEtran}
\usepackage{latexsym}
\usepackage{graphicx}
\usepackage{multirow}
\usepackage{algorithmic}
\usepackage{algorithm}
\usepackage{multirow}
\usepackage{color}
\usepackage{url}
\usepackage{umlaute}
\usepackage{amsmath}
\usepackage{amsthm}
\usepackage{eurosym}

\definecolor{red}{rgb}{1,0,0}

\ifCLASSINFOpdf
\else
\fi
\begin{document}
%
\title{Community Detection via Local Dynamic Interaction}




%
\author{\IEEEauthorblockN{Junming Shao\IEEEauthorrefmark{1},
Zhichao Han\IEEEauthorrefmark{1} and
Qinli Yang\IEEEauthorrefmark{1}}

\IEEEauthorblockA{\IEEEauthorrefmark{1,2,3}Web Sciences Center, University of Electronic Science and Technology of China, Chengdu, China\\
junmshao@uestc.edu.cn, hanchao0202@gmail.com, qinli.yang@uestc.edu.cn }

}


\maketitle

\begin{abstract}
\label{sec:abstract}
How can we uncover the natural communities in a real network that allows insight into its underlying structure and also potential functions? In this paper, we introduce a new community detection algorithm, called Attractor, which automatically spots the communities or groups in a network over time via local dynamic interaction. The basic idea is to envision a network as a dynamical system, and each agent interacts with its local partners. Instead of investigating the node dynamics, we actually examine the change of ``distances" among linked nodes. As time evolves, these distances will be shrunk or stretched gradually based on their topological structures. Finally all distances among linked nodes will converge into a stable pattern, and communities can be intuitively identified. Thanks to the dynamic viewpoint of community detection,  Attractor has several potential attractive properties: (a) Attractor provides an intuitive solution to analyze the community structure of a network, and faithfully captures the natural communities (with high quality). (b) Owing to its time complexity $O(|E|)$,  Attractor allows finding communities on large networks. (c) The small communities or anomalies, usually existing in real-world networks, can be well pinpointed. (d) Attractor is easy to parameterize, since there is no need to specify the number of clusters. Extensive experiments on synthetic and real-world networks further demonstrate the effectiveness and efficiency of the proposed approach.
\end{abstract}


%
\IEEEpeerreviewmaketitle

\section{Introduction}
\label{sec:intro}
During the past decades, community detection (also called graph clustering or graph partitioning)  has attracted a lot of attention. Many approaches have been proposed to identify communities based on different criteria, and each criterion (e.g. \emph{betweenness} \cite{Newman2002}, \emph{normalized cut}\cite{ncut},  \emph{modularity} \cite{Newman2006}, etc.) comes to specific advantages and drawbacks. Although many established approaches have already achieved some success, finding the intrinsic communities in complex networks is still a big challenge \cite{Evans2010}. As an example, the wide-spread modularity based algorithms \cite{Newman2006}, only yield a good graph partitioning if the network follows the random null assumption that each node has the equal chance to link any other node of the network \cite{Resolution}. This assumption becomes unreasonable for large networks (usually called ``resolution limit") as the connectivity pattern is usually in a local rather than a global fashion. Moreover, the growing size of networks in diverse fields is posing an increasing challenge for most established community detection algorithms.

In this paper,  instead of optimizing some user-defined criteria (e.g. \emph{normalized cut} or \emph{modularity}), we consider community detection from a new point of view: \emph{local distance dynamics}. We will see that the new viewpoint supplements an intuitive way to automatically identify community structure over time, and has several attractive properties. But let us first illustrate the basic idea.

\begin{figure*}[tb]
\centering
\includegraphics[height=40mm]{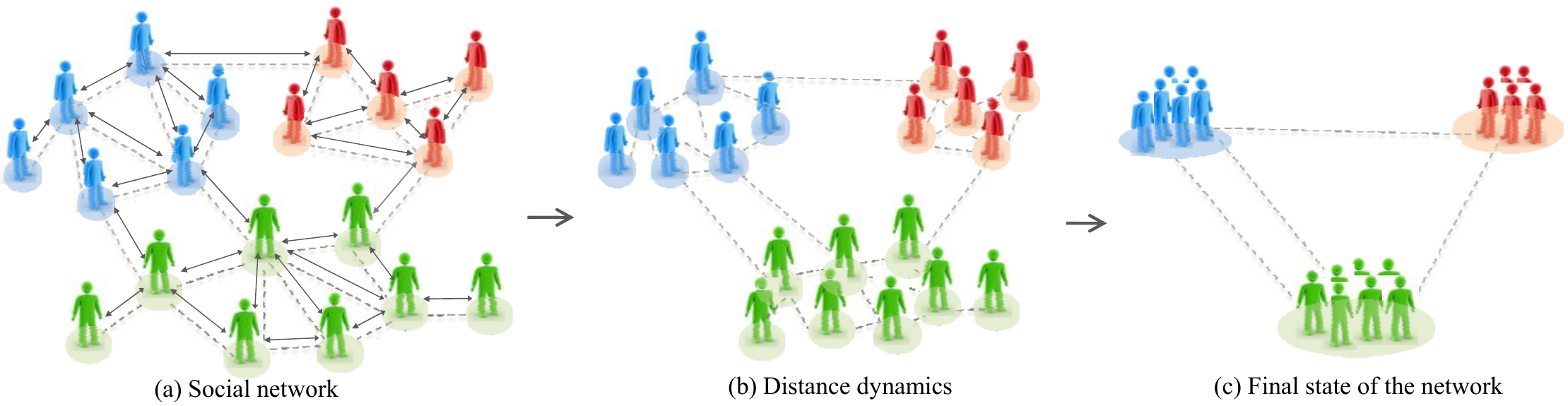}
\caption {Illustration of community detection via local dynamic interaction. (a) A social network, where the dashed lines indicate the relationships among persons, and arrows demonstrate the direct mutual interactions based on their relationships. (b) Relying on a proposed interaction model, the disparities of opinions among people will change over time,  where persons in the same community tend to gradually move together while people in different communities will keep far away from each other. (c) The final states of persons in terms of the ``distances": three intuitive communities.}
\label{fig:idea}
\end{figure*}

\subsection{Basic Idea}
\label{sec:idea}
In this paper, we expect to provide an intuitive way to shed light on the compartmental organization of a given network. The basic idea is to view a network as a dynamical system, and investigate its local dynamic interaction process over time. Instead of investigating the dynamics of nodes like in the traditional dynamical systems, we do capture the change of ``distances" among linked nodes over time. Thanks to the topology-driven interactions, the node distances will change gradually, where the change of node distances in the same community and between different communities will exhibit distinct dynamics over time, and finally all distances achieve a stable pattern. We call the stable distance pattern as an \emph{attractor}, a conceptual metaphor that all nodes try to attract their connected nodes, which results in the nodes sharing the same community move together while the nodes in different communities keep far way from each other. The new viewpoint has several benefits for community detection. Firstly, as a data representation, networks are characterized as a large number of interconnected units, such as persons with relationships, or interactions among proteins. Modeling the change of the relationships provides an intuitive image of the network dynamics. More importantly, insight into distances provides a more generalized way to analyze the networks in a metric space, rather than in a vector space in traditional dynamical analysis. This is quite beneficial to network analysis as the only information of real-world networks we usually can gain is their connectivity patterns.

Generally, the local dynamic interaction process involves three stages: First, the distances among linked nodes start with a set of initial values. As time evolves, since each node attracts its neighbors according to the proposed local interaction model (cf. Section \ref{sec:main}), the node distances in the same community tend to be gradually shrunk while those in different communities will be stretched.  Finally all distances will converge, and the network will be naturally split into several distinct communities by simply removing the edges associated with maximal distance (i.e. all distances equivalent to 1). To better illustrate the basic idea, let us take a social network as an example.  Fig.~\ref{fig:idea} displays the dynamics of an artificial social network composing of a number of persons and a set of inter-relationships (dashed lines). In this network, there exist three groups (representing as cartoon people with different colors) based on their different hobbies. Supposing some new techniques have been plugged into a mobile phone, and persons in this network are discussing their cons and pros. We are interested in knowing how opinion disparities among persons evolve driven by the underlying structure over time. In the beginning, each person usually has their own ideas, and the disparities of opinions with his neighbors are thus different (Fig.~\ref{fig:idea}(a)). Due to the influence from his/her known persons (i.e. persons having relationships),  the disparities of opinions among these persons will gradually change (increase or decrease) over time (Fig.~\ref{fig:idea}(b)). Finally the opinion disparities of all people tend to converge, and three communities are naturally popped up in terms of the ``distances" among persons (Fig.~\ref{fig:idea}(c)).

\subsection{Contributions}
\label{sec:contributions}
By investigating the dynamics of node distances with a local dynamic interaction process,  Attractor has several attractive benefits for community detection in networks, most importantly:
\begin{enumerate}
\label{sec:keypoints}
\item  \textbf{Intuitive Community Detection}: Instead of optimizing user-specified measures, Attractor investigates the community structure in networks from a local distance dynamics point of view. Building upon three interaction patterns, an intuitive interaction model has been introduced to explore the change of distances among linked nodes over time. The dynamics of node distances in the same communities and those between different communities exhibit different behaviors, and finally Attractor allows automatically spotting the communities in a network intuitively.

\item \textbf{Small Community and Anomaly Detection}: Driven by the local dynamic interactions, the small communities or anomalies usually existing in large-scale networks can be well identified.

\item \textbf{Scalability}: Thanks to the local interaction model, Attractor only needs to investigate the distances of linked nodes over time, which results in a relatively low time complexity of $O(|E|)$. This property of Attractor lends itself to handling large real-world networks.

\item \textbf{Parametrization}:  Attractor is easy to parameterize, as there is no need to specify the desirable number of clusters, instead of a cohesion parameter which is intuitive and robust to clustering results.
\end{enumerate}

The remainder of this paper is organized as follows: In the following section, we briefly survey related work. Section \ref{sec:main} presents our algorithm in detail. Section \ref{sec:experiment} contains an experimental evaluation. We finally conclude the paper in Section \ref{sec:conclusion}.

\section{Related Work}
\label{sec:relWork}
During the past several decades, many approaches have been established for community detection, such as \cite{Karypis98afast}, \cite{ncut}, \cite{Newman2006} etc. Due to space limitations, we only report the closest approaches from the literature. For detailed reviews of graph clustering, please refer to \cite{Schaeffer2007}.

\textbf{Cut-Criterion Clustering.} The cut-criterion based community detection algorithms refers to a class of widely used techniques which seek to partition a graph into disjoint subgraphs such that the number of ``cuts" across the subgraphs is minimized. Wu and Leahy  \cite{mincut} have proposed a clustering method based on the minimum-cut criterion, where the cut between two subgraphs is computed as the total weights of the edges that have been removed. $k-$disjoint subgraphs are obtained by recursively finding the minimum cuts that bisect the existing segments. To avoid an unnatural bias towards splitting small-sized subgraphs based on the minimum-cut criterion,  Shi and Malik \cite{ncut} have proposed the popular \emph{normalized cut}, to compute the cut cost as a fraction of the total edge connections to all the nodes in a graph. To optimize this criterion, a generalized eigenvalue decomposition was used to speed up computation time. In many cases, this class of graph clustering algorithms relying on the eigenvector decomposition of a similarity matrix is also called spectral clustering. Recently, modularity-based criterion has been introduced to measure the division of a network into communities.  Modularity-based graph clustering methods \cite{Newman2004}, \cite{Newman2006} partition a network into groups to ensure the number of edges between two groups is significantly less than the expected edges (i.e. ``expected cut").

\textbf{Multi-Level Clustering.}
Metis is a class of multi-level scalable partitioning techniques proposed by Karypis and Kumar \cite{multilevel}, \cite{Karypis98afast}. Graph clustering starts with constructing a sequence of successively smaller (coarser) graphs, and a bisection of the coarsest graph is applied. Subsequently, a finer graph is generated in the next level based on the previous bisections. At each level, an iterative refinement algorithm such as Kernighan-Lin (KL) or Fiduccia-Mattheyses (FM) is used to further improve the bisection. A more robust overall multilevel paradigm has been introduced by Karypis and Kumar \cite{Karypis98afast}, which presents a powerful graph coarsening scheme. It uses simplified variants of KL and FM to speed up the refinement without compromising the overall quality. Thanks to the multi-level graph construction, Metis also allows scaling up very large-scale networks.

\textbf{Markov Clustering.}
The Markov Cluster algorithm (MCL)  \cite{mcl} is a popular algorithm used in life sciences based on the simulation of (stochastic) flow in graphs. The basic idea is that dense regions in sparse graphs correspond to regions in which the number of random walks of length $k$ is relatively large. MCL basically identifies high-flowing regions representing the graph clusters by using an inflation parameter to separate regions of weak and strong flow.

\section{Community Detection by Local Dynamic Interaction}
\label{sec:main}
In this section, we present a new community detection approach based on local dynamic interaction.  The basic philosophy is to envision a network as a dynamic system, and dynamically investigate the distances among linked nodes to uncover its community structure. In the following, we start with some preliminary definitions, and then an interaction model is proposed in Section \ref{sec:model}. In Section \ref{sec:alg} we discuss the algorithm Attractor in detail, and analyze its time complexity in Section \ref{sec:time}.

\subsection{Preliminaries}
\label{sec:preli}
For the purpose of community detection, some necessary definitions are first introduced. \vspace{1mm}

\hspace{-3mm}\textsc{\textbf{Definition 1}} \hspace{1mm} (\textsc{Undirected Graph })
 Let $G=(V,E,W)$ be an undirected graph, where $V$ is the set of nodes, $E$ is the set of edges and $W$ is the corresponding set of weights. $ e = \{u,v\} \in E$ indicates a connection between the nodes $u$ and $v$.  $w(u,v)$ represents the weight of edge $e$.  $\forall e = \{u,v\} \in E, w(u,v) = 1$, in case of unweighted graph.

\vspace{1mm}
\hspace{-3mm}\textsc{\textbf{Definition 2}} \hspace{1mm} (\textsc{Neighbors of node $u$})
Given an undirected graph $G=(V,E,W)$, the neighborhood of a node $u \in V$ is the set $\Gamma(u)$ containing node $u$ and its adjacent nodes.
 \begin{equation}
\Gamma(u) =\{v \in V | \{u,v\} \in E\} \cup \{u\}
\label{eq:nb}
\end{equation}

Based on the two definitions, the similarity between any two nodes is further defined.  In this study,  we use the popular Jaccard distance \cite{jaccard} to quantify the node distance.  Selecting this measure mainly has two reasons. First, Jaccard distance provides an intuitive way to characterize the node distance. Generally, the more common neighbors two nodes have, the more similar they are. Secondly, Jaccard distance is computed in a local fashion and is thus time efficient. \vspace{2mm}

\hspace{-3mm}\textsc{\textbf{Definition 3}} \hspace{1mm} (\textsc{Jaccard Distance})
Given an undirected graph $G=(V,E,W)$,  the Jaccard distance of two nodes $u$ and $v$ is defined as:
\begin{equation}
d(u,v) = 1- \frac{|\Gamma(u) \cap \Gamma(v) |}{|\Gamma(u) \cup \Gamma(v) |}
\label{eq:jaccard}
\end{equation}

 For weighted graph, the Jaccard distance of two nodes $u$ and $v$ is further extended as:
 \begin{equation}
d(u,v) = 1- \frac{\sum_{x\in \Gamma(u) \cap \Gamma(v)}(w(u,x) +w(v,x))}{\sum_{\{x,y\}\in E; x,y \in \Gamma(u) \cup \Gamma(v)}  w(x,y)}
\label{eq:wjaccard}
\end{equation}

 Based on the definitions, we formally give the problem of community detection as follows.\vspace{1mm}

 \hspace{-3mm}\textsc{\textbf{Definition 4}} \hspace{1mm} (\textsc{Community Detection})
Given a graph $G=(V,E,W)$,  the problem of community detection is to partition a network into any $k$ disjoint communities $C_1,\cdots,C_k$ such that $V= C_1\cup,\cdots,\cup C_k$, and $C_i \cap C_j = \emptyset$.

 \begin{figure*}[tb]
\centering
\includegraphics[height=38mm]{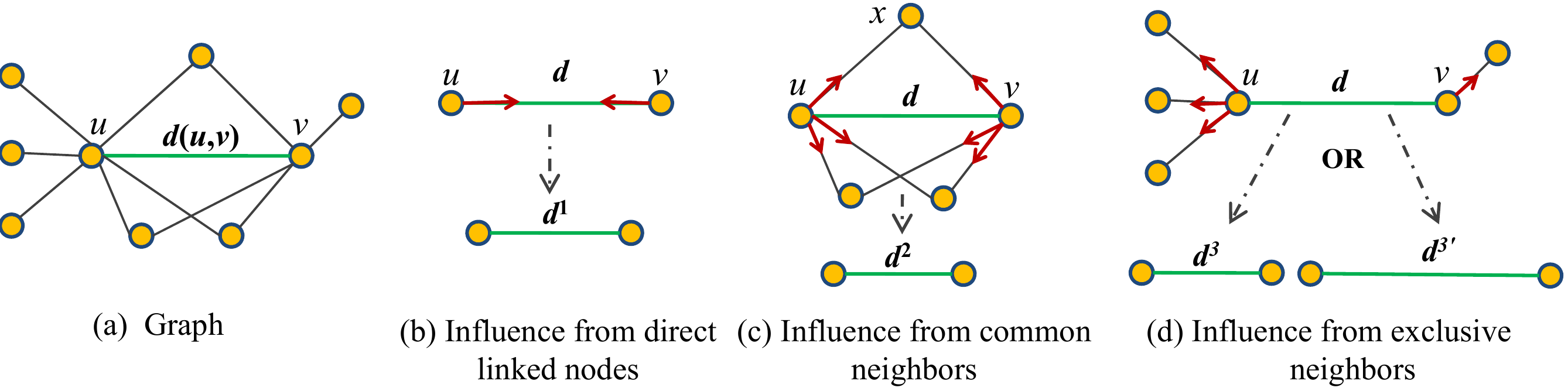}
\caption {Illustration of the change of node distances influencing from three distinct interaction patterns.}
\label{fig:patterns}
\end{figure*}

\subsection{Local Interaction Model}
\label{sec:model}
From the view of sociology,  a ``community" can be perceived as a group or network of persons who are connected to each other by relatively durable social relations to form a tight and cohesive social entity, due to the presence of a ``unity of will" or sharing common values \cite{Paul}. Motivated by such perception, it is curious for us to know whether the community structure can be automatically revealed by investigating the degree of cohesiveness of persons over time. Namely, we expect all persons in the same community will gradually enhance the cohesiveness by influencing each other, and finally converge together (e.g. same opinion, common values, etc.). In contrast, people in different communities will keep far away from each other and the community structure can be popped up intuitively. Therefore, in this study we try to reveal the community structure by investigating the dynamics of node distances over time based on an interaction model.

However, in order to build up an interaction model, we should answer the two following questions separately.

\vspace{2mm}
\textbf{ Q1:} How to determine a suitable interaction range for each node, in a local or global way?

\hspace{1mm}\textbf{Q2:}  How to determine the interaction patterns to provide a more natural interaction model?

\vspace{2mm}
\textbf{Interaction Range}.  In order to exploit the communities of a network, the local structure of network should be investigated. Therefore, instead of observing the collective interactions of all nodes,  we focus on the interaction in a local way. Namely, for each node, it interacts with its local neighbors rather than all other nodes spreading in the network.  Obviously, the intrinsic connectivity pattern of real-world networks gives a natural way to model the interaction range.

\textbf{Interaction Patterns}. After specifying the local interaction range, the next crucial step is to determine the interaction patterns among nodes. As we expect to analyze the community structure in networks by investigating the node distances, the interaction patterns and the corresponding influence on node distances should be clarified. Formally, let $ e = \{u,v\} \in E$ be an edge between two linked nodes $u$ and $v$, and $d(u,v)$ is its initial distance. Obviously, the change of $d(u,v)$ results from the variation of $u$ and $v$ (e.g. opinion disparity is caused by the opinion change of two persons due to mutual discussion). Hence, any interaction resulting in the change of the corresponding two nodes (i.e. $u$ and $v$) needs to be considered to investigate the influence on $d(u,v)$. Generally, regarding any edge between two adjacent nodes, there are three distinct interaction scenarios influencing its distance relying on its topological structure.

\begin{figure*}[tb]
\centering
\includegraphics[height=40mm]{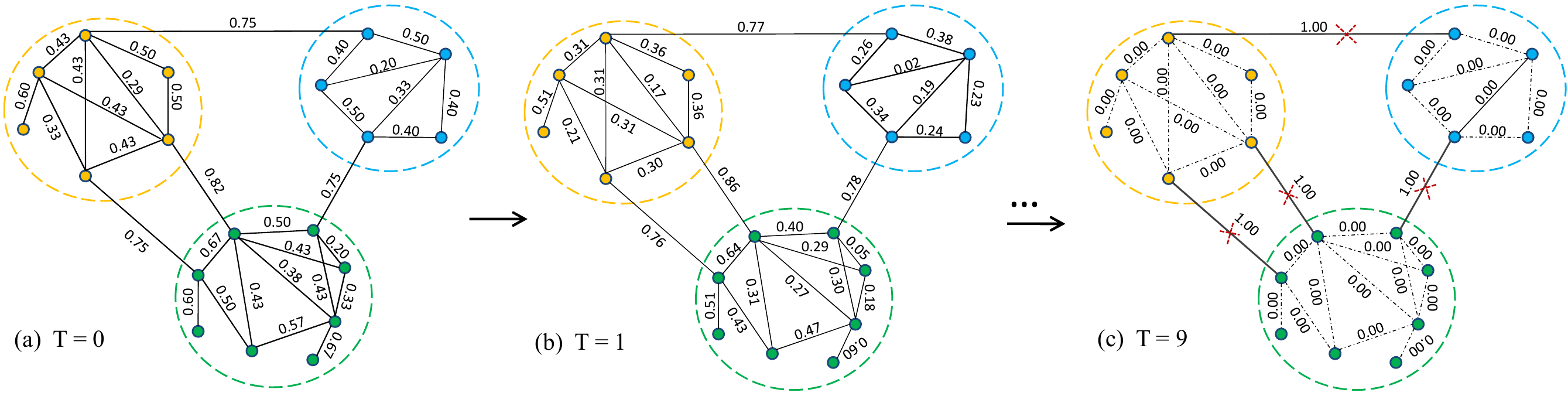}
\caption {Illustration of the distance dynamics. (a) The graph representation of the social network of Fig. 1(a), where the numbers on edges indicate the initial distances among connected nodes. (b) The updated node distances after one time step. (c) The final state of the network.}
\label{fig:dynamics}
\end{figure*}

\vspace{2mm}
\textsc{Pattern 1: Influence from direct linked nodes}  As node $u$ and node $v$ are linked,  each node attracts each other, and makes the opposite node move to itself. The distance between node $u$ and $v$ is thus shrunk. Like a friendship network, each people affects their known people, and tends to increase their cohesiveness gradually. Formally, we characterize the change of  $d(u,v)$ from the interaction of direct linked nodes, $DI$, as follows:

\vspace{-1mm}
\begin{equation}
DI= -\bigg{(}\frac{f(1-d(u,v))}{deg(u)} + \frac{f(1-d(u,v))}{deg(v)}\bigg{)}
\label{eq:pattern1}
\end{equation}

where $deg(u)$ is the degree of the node $u$, $f(\cdot)$ is a coupling function and $sin(\cdot)$ is used in this study. $1-d(u,v)$ indicates the similarity between $u$ and $v$, and the more similar the two nodes are, the higher influence between each other they will have. The term $\frac{1}{deg(u)}$ is called normalization factor, which is used to consider the different influences between linked nodes with diverse degrees. Namely, the nodes with more links are harder to be influenced comparing to the nodes with less links. Take instructor network as an example that one supervisor usually links to many students while one student only connects to his supervisor. In this situation, the supervisor may have a high influence on each student while the influence for supervisor from each student is relatively low. For illustration, Fig. \ref{fig:patterns} gives an example, and the change of the distance from direct linked nodes is demonstrated in Fig. \ref{fig:patterns}(b).

\vspace{2mm}
\textsc{Pattern 2: Influence from common neighbors} Here we consider the second scenario: the influence from the common neighbors $CN = (\Gamma(u) - u) \cap (\Gamma(v) - v)$ of nodes $u$ and $v$ (Fig. \ref{fig:patterns}(c)). As the common neighbors have both links with the two nodes $u$ and $v$,  they attract the two nodes and will result in the change of the distance $d(u,v)$. Namely, the common neighbors attract both node $u$ and node $v$ to move to them, and lead to the decrease of the distance $d(u,v)$ (See Fig. \ref{fig:patterns}(c)). Formally, we define the change of $d(u,v)$ from the influence of common neighbors, $CI$, as follows:

\vspace{-1mm}
\begin{displaymath}
CI=-\sum_{x \in CN} \bigg{(}\frac{1}{deg(u)}\cdot f\big{(}1-d(x,u)\big{)} \cdot (1-d(x,v)) +
\end{displaymath}\vspace{-2 mm}
\begin{equation}
\hspace*{12mm}
\frac{1}{deg(v)}\cdot f\big{(}1-d(x,v)\big{)}\cdot (1-d(x,u))\bigg{)}
\label{eq:pattern2}
\vspace{-1mm}
\end{equation}
Here the two terms $(1-d(x,v))$ and $(1-d(x,u))$ for each common neighbor are used to further quantify the degree of influence compared to the influence from direct linked nodes. For example, considering a common neighbor $x$ interacting with node $u$ (see Fig. \ref{fig:patterns}(c)),  if $x$ and $v$ are more similar, the influence from $x$ on $u$ is more similar to the influence from $v$. Theoretically, once the similarity between $x$ and $v$ equals one (i.e. they can be viewed as the same node), the influence of the node $x$ on the distance $d(u,v)$ simply transfers into the first pattern.

\vspace{2mm}
\textsc{Pattern 3: Influence from exclusive neighbors}: The third interaction pattern happens when there exists some neighbors exclusively belong to node $u$ or $v$,  $EN(u) =\Gamma(u) - \Gamma(u) \cap \Gamma(v)$ , $EN(v) =\Gamma(v) - \Gamma(u) \cap \Gamma(v)$, respectively. Although, like pattern 1 and pattern 2, each exclusive neighbor of $u$ attracts $u$ to move close to itself, there is no knowledge whether node $u$ is attracted to move closer to node $v$ or attracted to move far away from $v$  (see Fig. \ref{fig:patterns}(d)). To determine the positive or negative influence of exclusive neighbors on the distance, a similarity-based heuristic strategy is proposed. The basic philosophy is to investigate whether each exclusive neighbor of node $u$ is similar with node $v$, and vice versa. If the exclusive neighbor of node $u$ is similar with node $v$, the movement of node $u$ towards exclusive neighbor results in the decrease of the distance $d(u,v)$. Similarly, If the exclusive neighbor is not similar with node $v$, the movement of node $u$ towards the exclusive neighbor will lead to the opposite effect: moving far away from the node $v$.  Therefore, here we introduce a cohesion parameter $\lambda$, to determine the underlying influence as follows. The cohesion parameter $\lambda$ will be further discussed in Section \ref{sec:alg}.

\vspace{-3mm}
\begin{equation}
\rho(x,u) = \left\{ \begin{array}{ll}
\big{(}1-d(x,v)\big{)} & \textrm{$\big{(}1-d(x,v)\big{)}  \geq \lambda$ }\\
\big{(}1-d(x,v)\big{)} - \lambda  & \textrm{otherwise}
\end{array} \right.
\label{eq:sign}
\end{equation}

where $\rho(x,u)$ characterizes the degree of positive or negative influence on the distance $d(u,v)$. Then, the change of $d(u,v)$ influencing by exclusive neighbors, $EI$, is defined as follows:

\vspace{-3mm}
\begin{displaymath}
EI =-\sum_{x \in EN(u)} \bigg{(}\frac{1}{deg(u)}\cdot f\big{(}1-d(x,u)\big{)}\cdot \rho(x,u) \bigg{)}
\end{displaymath}\vspace{-2 mm}
\begin{equation}
\hspace*{12mm}
-\sum_{y \in EN(v)} \bigg{(}\frac{1}{deg(v)}\cdot f\big{(}1-d(y,v)\big{)}\cdot \rho(y,v) \bigg{)}
\label{eq:pattern3}
\end{equation}

Finally, by considering the three interaction patterns together,  the dynamics of  the distance $d(u,v)$ between nodes $u$ and $v$ over time is provided by:

\vspace{-3mm}
\begin{equation}
d(u,v,t+1) = d(u,v,t) + DI(t) + CI(t) + EI(t)
\label{eq:dynamics}
\end{equation}

where $d(u,v,t+1)$ is the renewed distance at time step $t+1$. $DI(t)$, $CI(t)$ and $EI(t)$ characterize the change of distance from the direct linked nodes, common neighbors and exclusive neighbors, respectively.

\begin{figure*}[tb]
\centering
\begin{tabular}{ccccc}
\includegraphics[height=32mm]{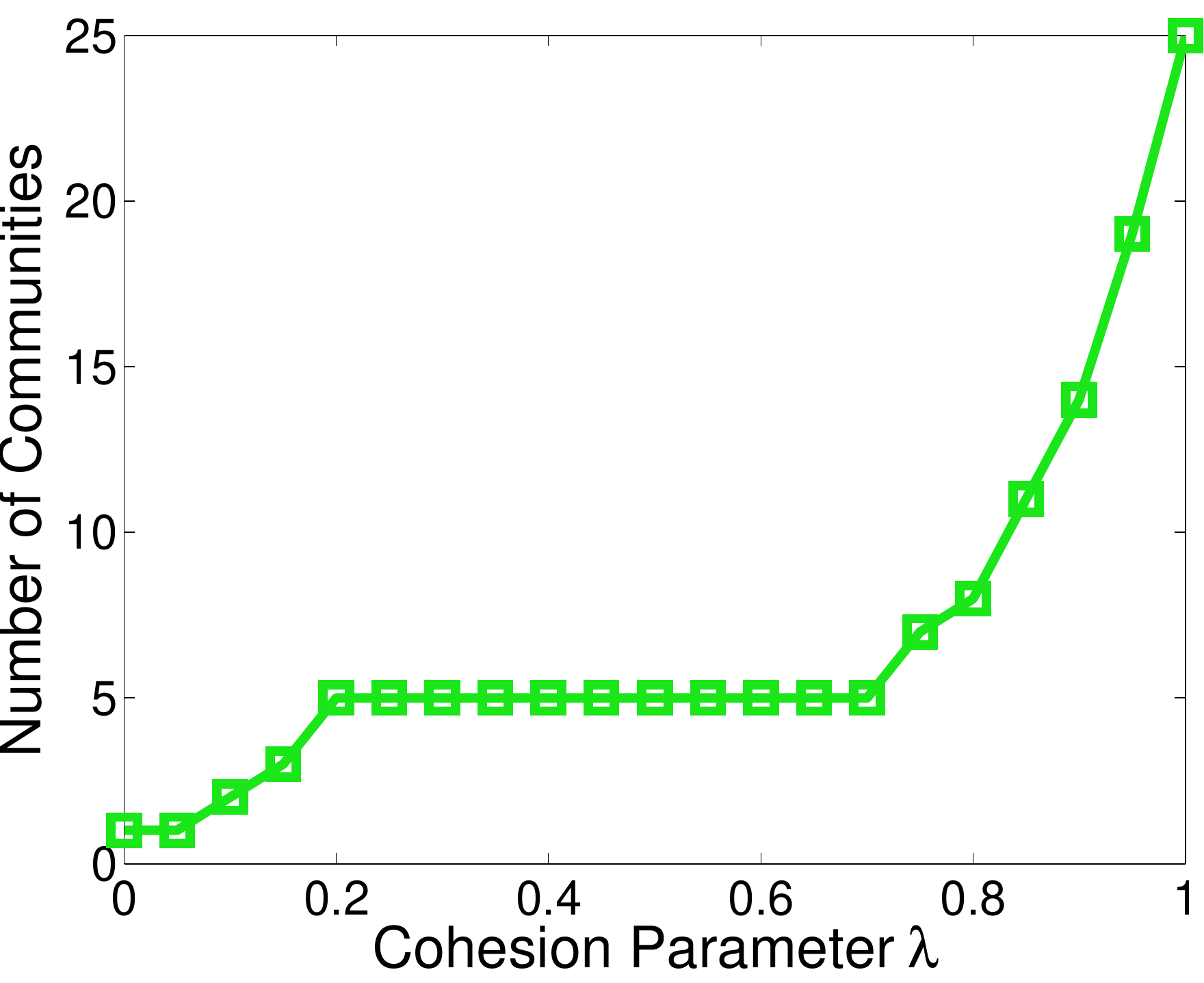}&
\includegraphics[height=32mm]{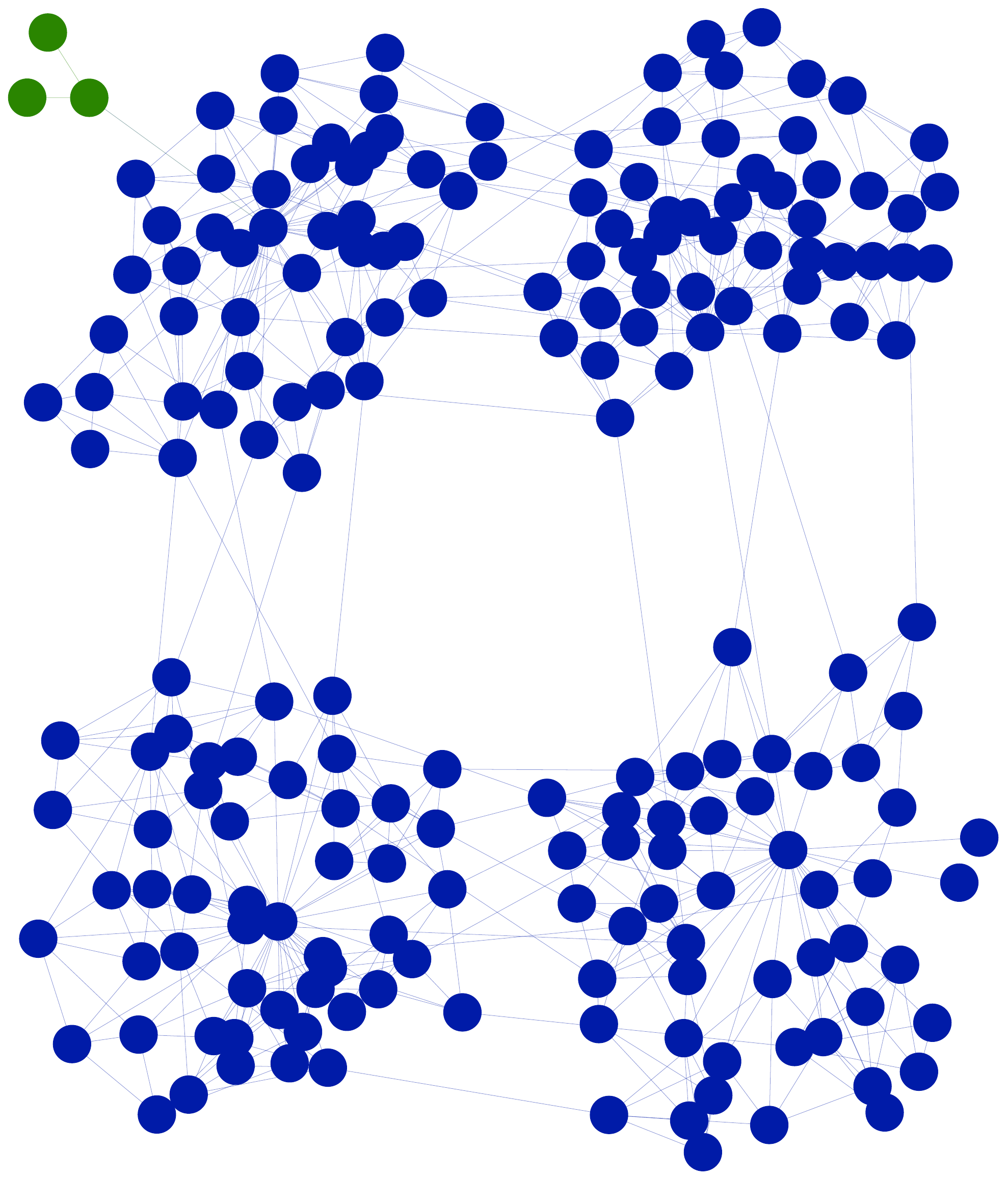}&
\includegraphics[height=32mm]{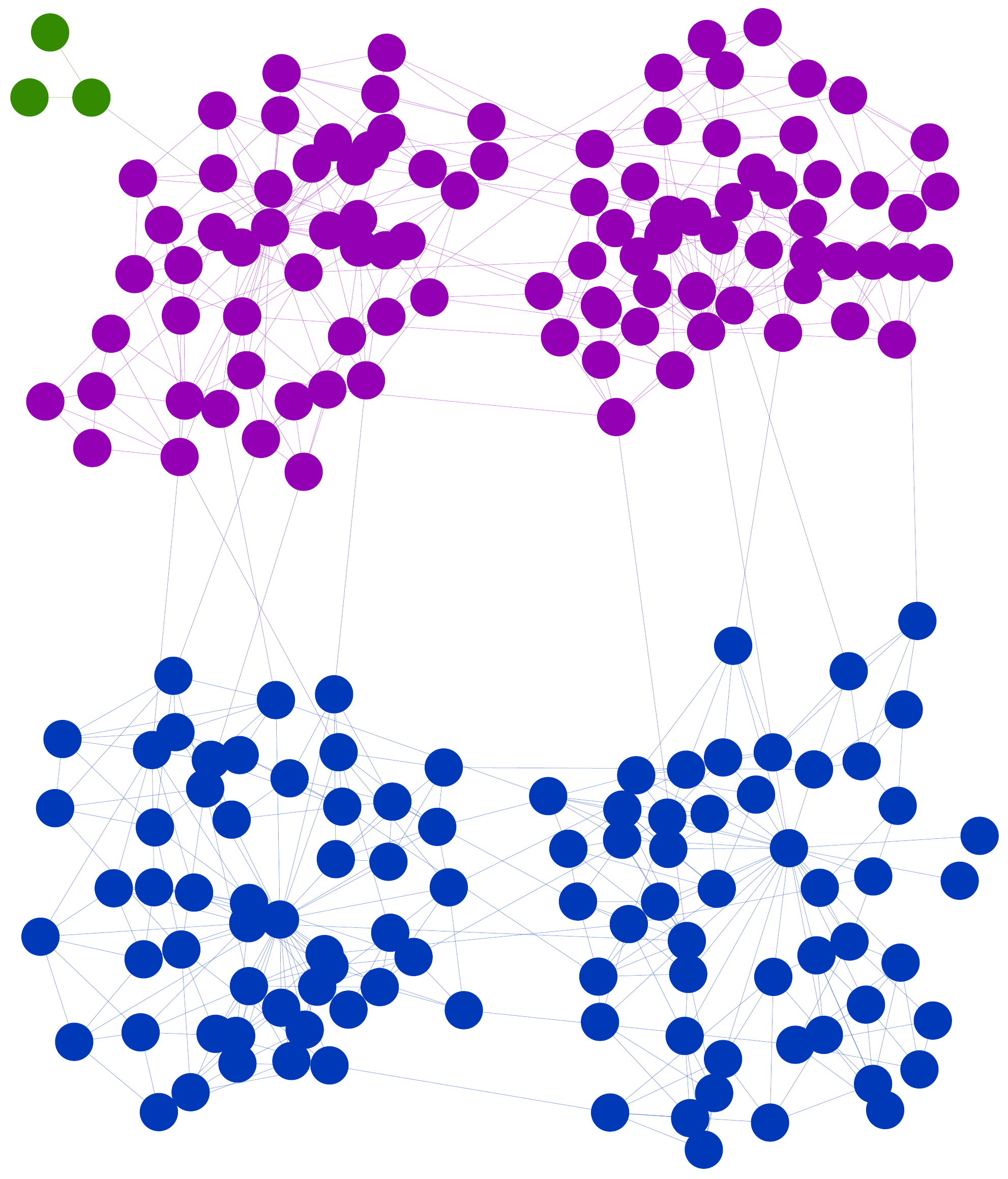}&
\includegraphics[height=32mm]{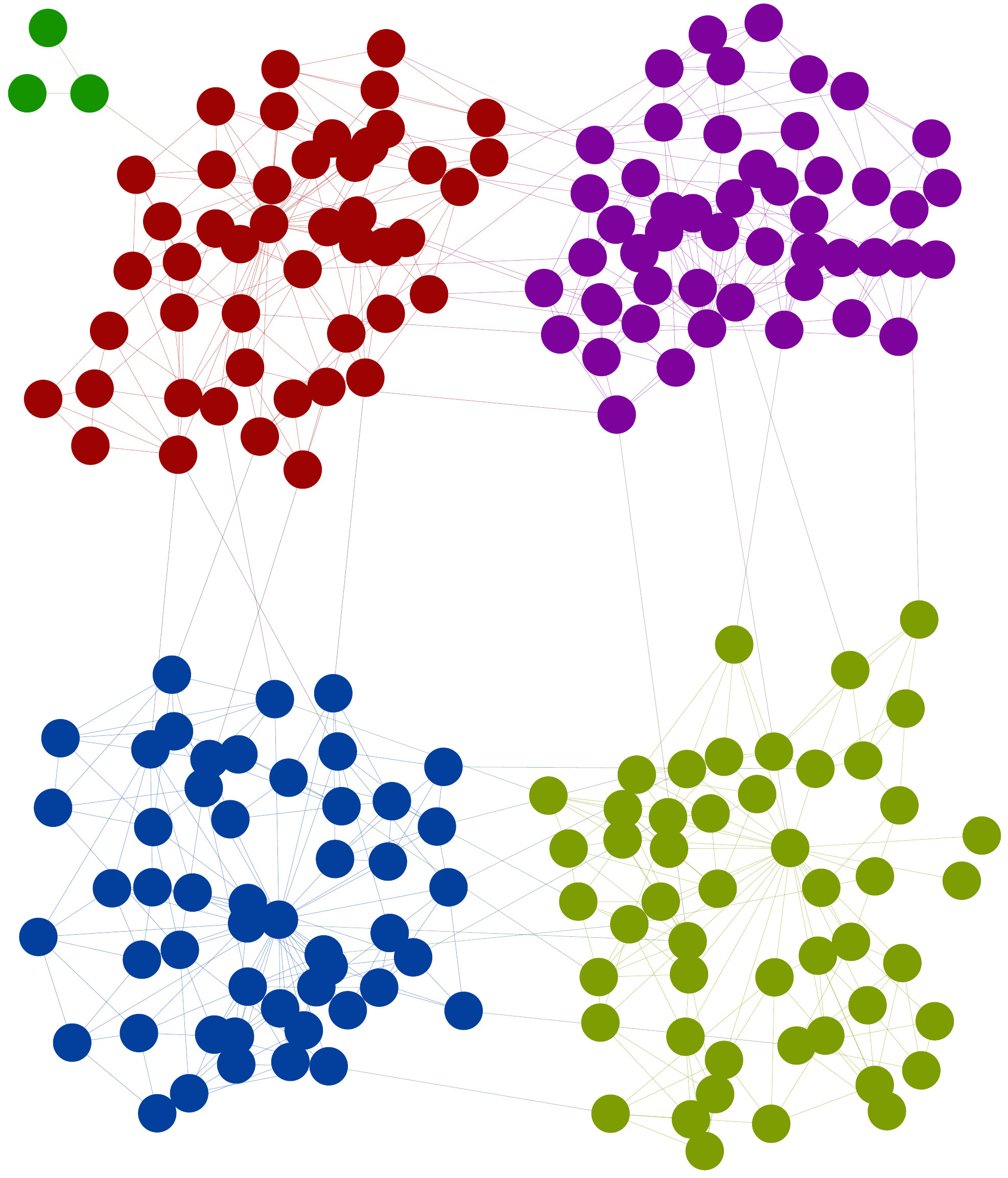}&
\includegraphics[height=32mm]{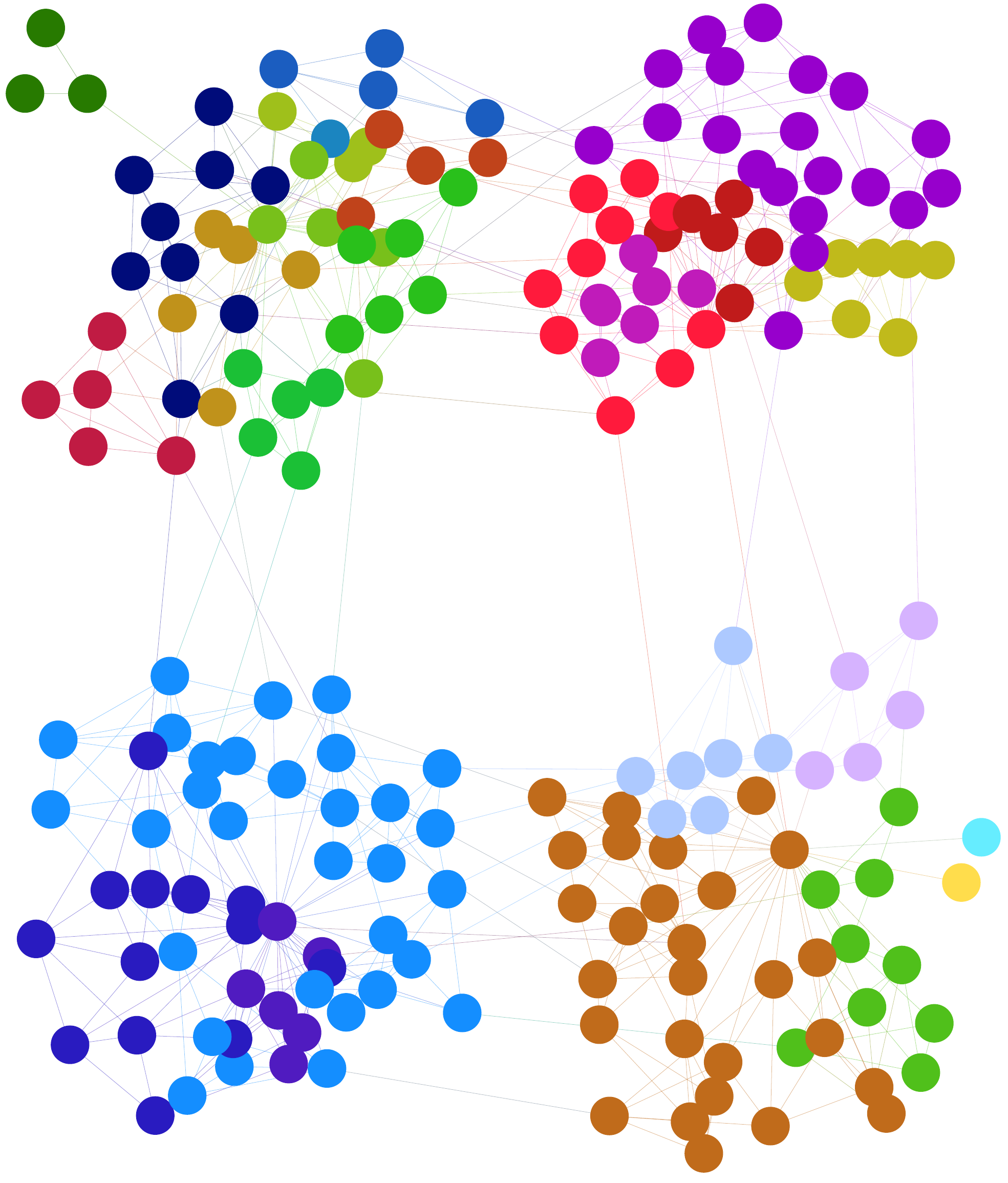}
\\
(a) $\lambda$ vs \#C & (b) $\lambda = 0.1$ &(c) $\lambda = 0.15$ &(d) $\lambda \in$ [0.2 - 0.7]& (e) $\lambda = 1$
\end{tabular}
\caption {The sensitivity of cohesion parameter $\lambda$ on community detection.}
\label{fig:levels}
\end{figure*}

\subsection{The Attractor Algorithm}
\label{sec:alg}
In this section, we present the Attractor algorithm in detail based on the proposed interaction model.

\textbf{Dynamical Interaction.}
\label{sec:flat}
Building upon the interaction model (cf. Eq. (\ref{eq:dynamics})), the distance dynamics in a network can be simulated, which mainly involves the following steps:
\begin{enumerate}
\item At initial time ($t=0$), without any interaction, each edge is associated with an initial distance. Here, the initial value is computed according to the Jaccord distance with Definition \ref{eq:jaccard} and Definition \ref{eq:wjaccard}.
\item As time evolves, each node intuitively interacts with its linked nodes, and the influence of each interaction on the distance is further captured according to the three interaction pattern scenarios (Eq. (\ref{eq:pattern1}), Eq. (\ref{eq:pattern2}) and Eq. (\ref{eq:pattern3})), respectively. Through the different influences, the node distances sharing the same community tend to decrease while the distances among nodes in different communities will increase gradually.
\item Finally, all distances will converge, and the communities can be easily obtained by removing the edges with maximal distances (i.e. $d(u,v)$ =1).
\end{enumerate}

For illustration, Fig. \ref{fig:dynamics}(a)-(c) shows three states for the social network of Fig. \ref{fig:idea} from $t=0$ to $t=9$ during the local dynamic interaction process. $T=0$ indicates the initial distances among connected nodes (Fig. \ref{fig:dynamics}(a)). From that moment on, each node interacts with its neighbors and influences the corresponding distances based on the proposed interaction model (cf. Eq. \ref{eq:dynamics}), and the new node distances after one time step are further illustrated in Fig. \ref{fig:dynamics}(b). After nine steps, all distances converge, either 0 or 1, and three communities are naturally identified by cutting out all edges with distance equaling to 1.

\begin{algorithm}[!tb]
\caption{Attractor}
\label{alg:attractor}
\begin{algorithmic}[1]
\small{
\STATE \textbf{Input:} $G = (V,E,W)$, $\lambda$ \\[1ex]
\STATE // Initialization of distances
    \FOR{each edge $e =\{u,v\} \in E$}
        \STATE compute the initial distance $d_e^{0}$ using Eq. (\ref{eq:wjaccard});
        \FOR{each node $x \in EN(u)$}
            \STATE compute the distance $d_{ux}^{0}$ using Eq. (\ref{eq:wjaccard});
        \ENDFOR
        \FOR{each node $y \in EN(v)$}
            \STATE compute the distance $d_{vy}^{0}$ using Eq. (\ref{eq:wjaccard});
        \ENDFOR
    \ENDFOR\\[1ex]
\STATE // Dynamic Interaction
\STATE Flag = TRUE;
\WHILE{Flag}
        \STATE Flag = FALSE;
        \FOR{each edge $e =\{u,v\} \in E$}
            \IF{$0<d_e^t<1$}
                \STATE Compute $DI_e^t$, $CI_e^t$, $EI_e^t$  using Eq.  (\ref{eq:pattern1}), (\ref{eq:pattern2}),  (\ref{eq:pattern3});
                \STATE $\Delta d_e^t = DI_e^t + CI_e^t +EI_e^t$;
                \IF{$\Delta d_e^t\neq0$}
                    \STATE // compute the renewable distance over time
                    \STATE $d_e^{t+1}$ =$d_e^t$ + $\Delta d_e^t$;
                    \IF{$d_e^{t+1}>1$}
                        \STATE $d_e^{t+1}$ = 1;
                    \ENDIF
                    \IF{$d_e^{t+1}<0$}
                        \STATE $d_e^{t+1}$ = 0;
                    \ENDIF
                    \STATE Flag = TRUE;
                \ENDIF
            \ENDIF
        \ENDFOR
 \ENDWHILE\\[1ex]
  \STATE //Find communities
         \FOR{each edge $e =\{u,v\} \in E$}
            \IF{$d_e^{t+1} =1$}
                   \STATE remove the edge $e$ from the network;
            \ENDIF
         \ENDFOR
          \STATE find the resulting components (communities) $C$;\\[1ex]
\STATE \textbf{Output:} $C$;
}
\end{algorithmic}
\end{algorithm}

\textbf{Detection of small communities or Anomalies.}
In real-world networks, there usually exist many communities with various sizes. Especially in large-scale networks,  the size of a large fraction of communities is usually small \cite{small}. However, for many traditional community detection algorithms, such as Modularity or Ncut, they tend to partition the whole network into relatively equal-size groups with cluster size being no less than $\sqrt{n}$ ($n$ is the number of nodes in a network) \cite{small}, and fail to find small communities due to the problem called ``resolution limit" \cite{Resolution}. For attractor, as it naturally models the dynamics of node distances and does not rely on any user-defined criteria, it allows intuitively finding the small communities in networks.  Moreover, it also provides a promising way to handle anomalies/outliers. In this scenario,  anomalies are interpreted as the noisy nodes or unusual nodes isolated from all other nodes over time, and finally pop out automatically.

\textbf{Cohesion parameter $\lambda$.}
For the Attractor algorithm, the cohesion parameter $\lambda$ is used to determine the positive or negative interaction influence on the distances from exclusive neighbors.  Generally, with the higher value of $\lambda$,  it yields more communities while produces bigger communities with lower value of $\lambda$. By modulating the cohesion parameter $\lambda$, Attractor allows analyzing the community structure from coarse to fine. Moreover, $\lambda$ is informative and is easy to tune compared to the algorithms requiring to specify the number of clusters. Fig. \ref{fig:levels}(a) plots the finding number of communities with different $\lambda$ ranging from 0 to 1 on a synthetic network. From this plot, we can see that Attractor allows yielding perfect partitioning with the parameter $\lambda$ on a long stable range (0.2 - 0.7). The clustering results with respect to distinct parameters are further illustrated in Fig. \ref{fig:levels}(b) to Fig. \ref{fig:levels}(e).  Extensive experiments further demonstrate Attractor is not sensitive to clustering results and usually produces a good result within the range $\lambda = [0.4, 0.6]$. Finally, the Pseudocode of Attractor is given in Algorithm \ref{alg:attractor}.

\subsection{Time Complexity Analysis}
\label{sec:time}
To investigate the distance dynamics,  the initial distance of any two linked nodes in a network is required, and thus the time computation is $O(|E|)$. Moreover, for the local dynamic interaction, Attractor also needs to compute the corresponding jaccard distances for exclusive neighbors (Algorithm \ref{alg:attractor}(Line 5-10)). The time complexity is $O(k\cdot |E|)$, where $k$ is approximately the average number of exclusive neighbors of two linked nodes. During the local dynamic interaction process, as all distances have already existed, Attractor only needs to recall these distances at previous time without any distance computation, and thus the time complexity is $O(T\cdot |E|)$. Totally, the time complexity is $O(|E|+ k\cdot |E| +T\cdot|E|)$ , where $T$ is the number of time steps. In most cases, $T$ is small with $3 \leq T \leq 50$.

\section{Experiments}
\label{sec:experiment}
In this section, we evaluate our proposed algorithm Attractor on synthetic as well as real-world networks to demonstrate its benefits.

\textbf{Selection of comparison methods.} To extensively study the performance of Attractor, we compare it to representatives of various community detection paradigms: we select the \emph{normalized-cut} criterion based graph clustering method Ncut \cite{ncut} and the popular \emph{modularity}-based community detection algorithm by Newman \cite{Newman2006} (called as Modularity), the well-known multi-level graph partitioning algorithm Metis by Karypis and Kumar \cite{Karypis98afast} and the flow simulation based Markov Cluster algorithm (MCL) by Dongen \cite{mcl}. For all experiments, without further statement, Ncut and Metis specify the cluster number $K = |C|$, $|C|$ is the true number of classes of the network if the ground truth is available. MCL takes the default inflation parameter ($i = 2.0$) as suggested by authors  \cite{mcl}. We set the cohesion parameter $\lambda = 0.5$ for Attractor as default parameter. All experiments have been performed on a workstation with 3.4 GHz CPU and 32.0 GB RAM.

\textbf{Evaluation measures:} To compare different community detection algorithms with respect to effectiveness, we evaluate the clustering results in two ways. First, if class information is available for a network, the clustering performance is directly measured by three widely used evaluation measures: \emph{Normalized Mutual Information (NMI)} \cite{nmi}, \emph{Adjusted Rand Index (ARI)} \cite{ari} and \emph{Cluster Purity}.  For the networks without ground truth, we apply the popular internal evaluation measures \emph{modularity} \cite{Newman2006} and \emph{normalized cut (cut)} \cite{ncut} to evaluate the quality of obtained communities by different algorithms although they are somehow tailored for \emph{modularity}-criterion or \emph{cut}-criterion based algorithms.

\begin{figure}[!tb]
\centering
\includegraphics[height=48mm]{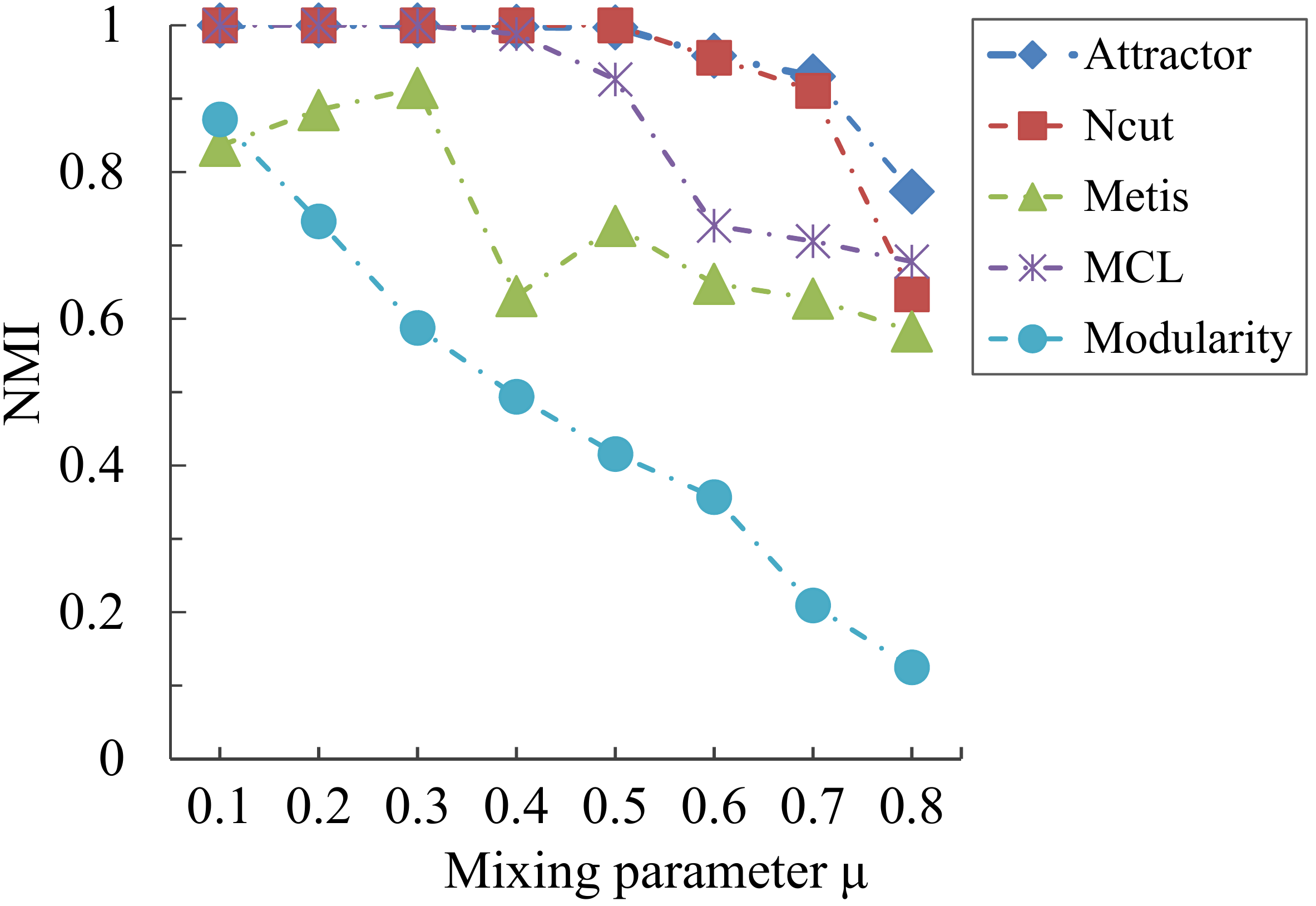}
\caption {The performance of different algorithms on the LFR benchmark networks by varying the number of inter-cluster edges using the mixing parameter $\mu$.}
\label{fig:noise}
\end{figure}

\subsection{Synthetic networks}
In this section, we first generate several synthetic networks featuring distinct characteristics to compare the performance of various community detection algorithms. For fair comparison and to make the synthetic networks to be more consistent with the real-world networks, the LFR benchmark networks \cite{benchmark} have been applied, where the distributions of degree and community size of networks can be easily tuned. To increase the complexity of networks, the \emph{mixing parameter} $\mu$ \cite{benchmark},  defined as the fraction of links of each node outside its community, is used to control the difficulty of community separation.

\begin{figure}[!tb]
\centering
\includegraphics[height=48mm]{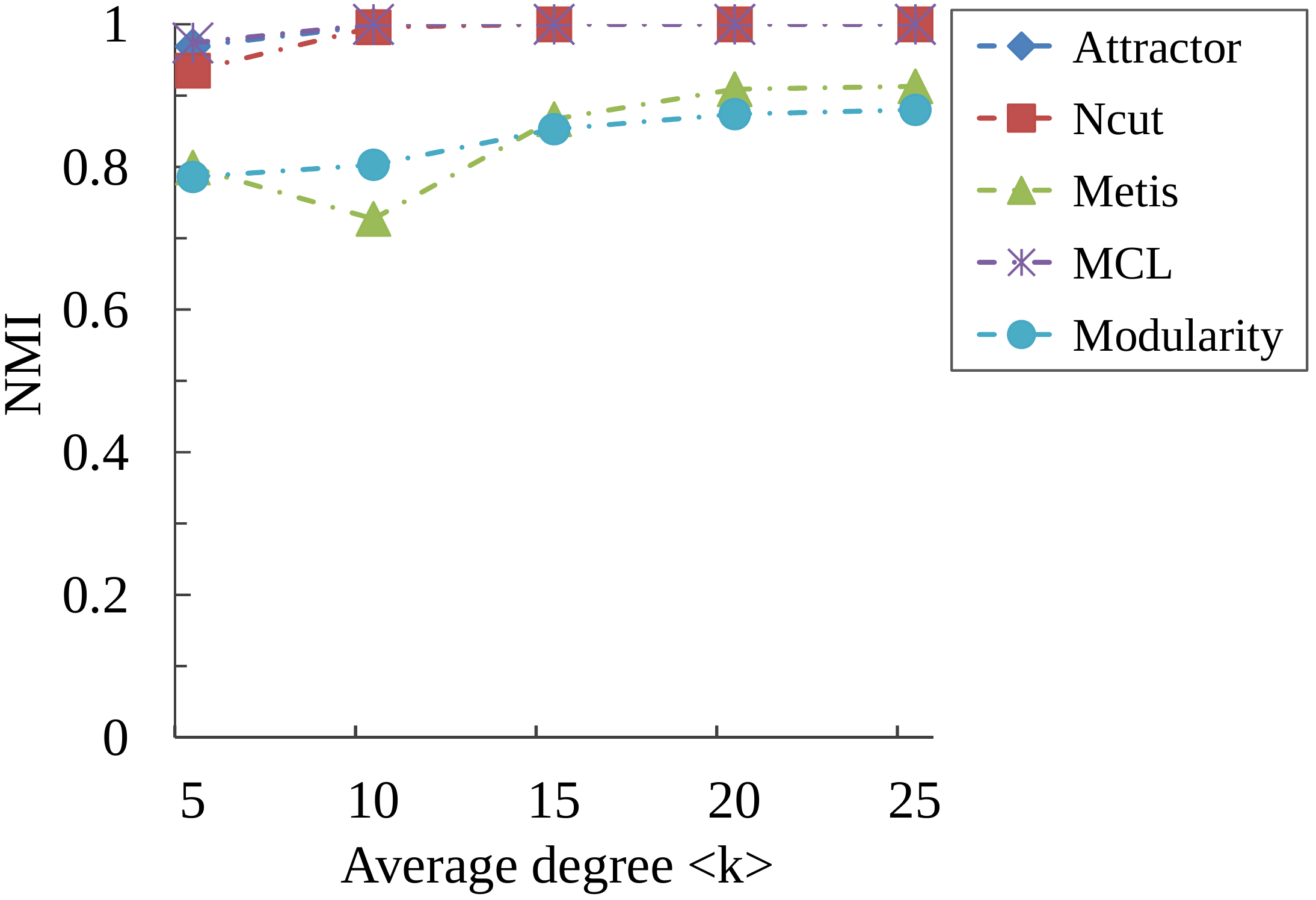}
\caption {Performance of different algorithms on the LFR benchmark networks by varying community density using the average degree $<$$k$$>$.}
\label{fig:density}
\end{figure}

%

\textbf{Noise Edge:} First, we evaluate how well the different graph clustering algorithms allow detecting communities by varying their inter-cluster edges. The inter-cluster edges,  which we call noise edges, are added into the network to hamper community separation. We fix node average degree and community size, and change the \emph{mixing parameter} $\mu$ from $0.1$ to $0.8$ to generate a serial of networks with different inter-cluster edges.  All networks consist of 2000 nodes with the average degree $k =20$.

With the increase of \emph{mixing parameter}, the performance (measured by NMI) of all five approaches is shown in Fig. \ref{fig:noise}. We can see that the algorithms of Attractor, Ncut and MCL almost achieve the perfect clusterings by adding inter-cluster edges with the \emph{mixing parameter} up to $0.4$ (40\% edges of each node links to other communities). The performance begins to decrease with more and more inter-edges added into the network, and Attractor is more robust to these noise edges. For Modularity,  it is more sensitive to these noise edges, and its performance is not comparable with other three algorithms on these networks. Regarding the Metis algorithm, the performance is fluctuated and starts to decrease dramatically as soon as more inter-edges are added (with $\mu = 0.4$).

 \textbf{Community Density:} Next, we evaluate how the algorithms respond to the networks with different average degrees, which we call community density.  Here we fix the inter-cluster edges ($\mu = 0.1$), and change the average degree $k$ from $25$ to $5$ to see the influence of community density on the performance of these algorithms.

Fig. \ref{fig:density} shows that Attractor, MCL and Ncut yield good results for all these networks, while the performances of Metis and Modularity are much worse. We can see that Attractor, Metis and Ncut allows correctly finding the good communities even with low community density ($k = 5$). For Metis and Modularity, they are more sensitive to the community density on these synthetic networks.

\begin{figure}[tb]
\centering
\begin{tabular}{cccc}
\includegraphics[height=32mm]{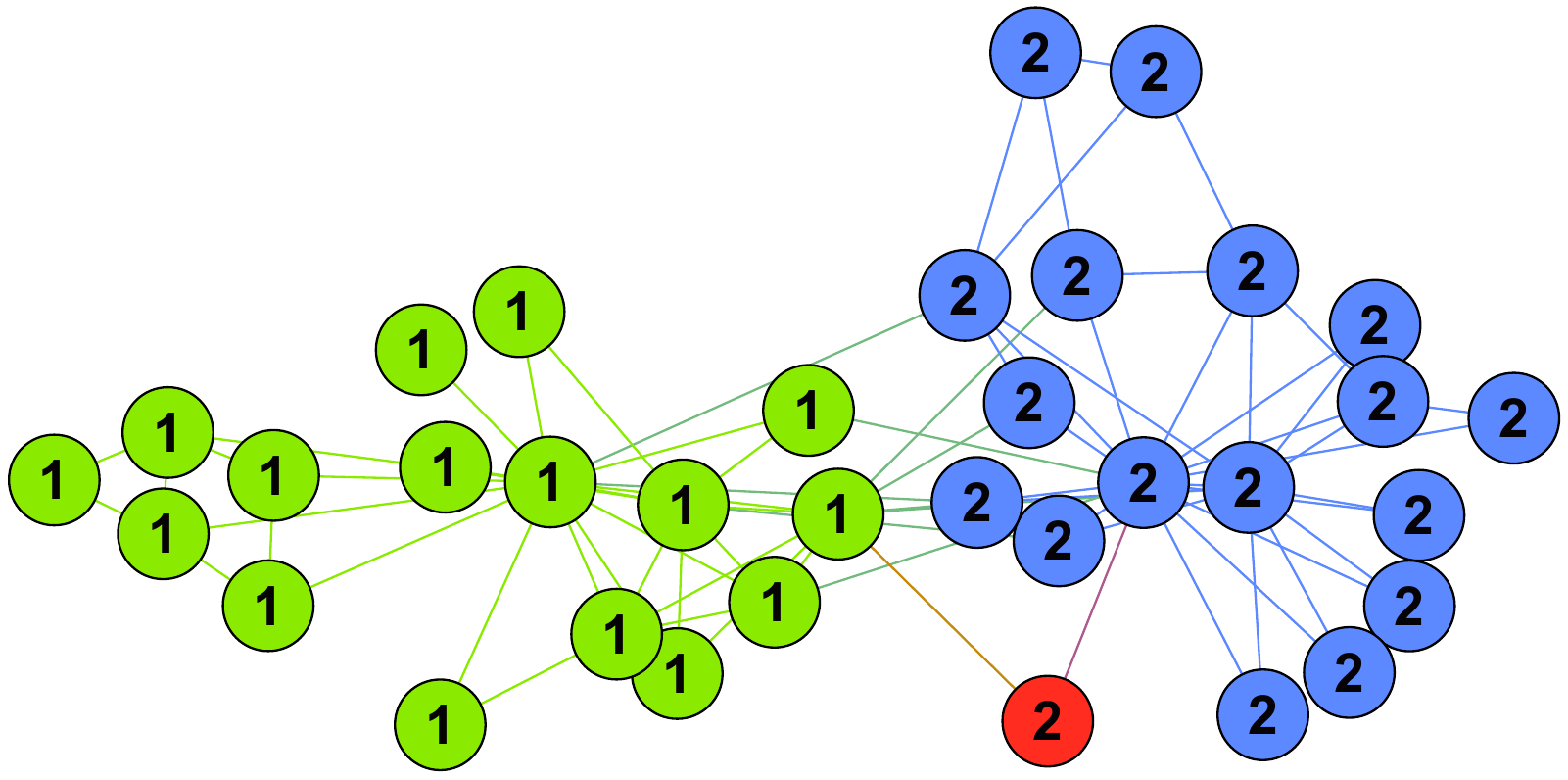}
\end{tabular}
\caption{Performance of Attractor on the karate club network, where the colors of nodes indicate different detected communities.\vspace{-3mm}}
\label{fig:karate}
\end{figure}

\tabcolsep=4pt
\begin{table*}[!htb]
\center \caption{Performance of different graph clustering algorithms on labeled real-world networks.}
\begin{tabular}{|c|c|c|c|c|c|c|c|c|c|c|c|c|c|c|c|c|c|c|c|c|c|c|c|c|c|}
\hline \multirow{2}{*}{Labeled Networks}  &  \multirow{2}{*}{$|V|$} &  \multirow{2}{*}{$|E|$} & \multicolumn{3}{|c|}{Attractor} & \multicolumn{3}{|c|}{Ncut} & \multicolumn{3}{|c|}{Modularity}& \multicolumn{3}{|c|}{Metis} & \multicolumn{3}{|c|}{MCL}\\ \cline{4-18}
   & & & NMI & ARI & Pur & NMI & ARI & Pur& NMI & ARI & Pur & NMI & ARI & Pur & NMI & ARI & Pur\\ \hline
Zarachy &34 &78 &0.859 &0.939 &1
&0.833 &0.882 &0.970
&0.577 &0.680 &0.970
&0.836 &0.882 &0.970
&0.833 &0.882 &0.970\\ \hline
College football & 115 & 613&0.923  & 0.897 &0.930
& 0.923 &0.897  &0.930
&0.596    &0.474&0.574
&0.393 &0.095 &0.339
&0.923 &0.897 &0.930\\ \hline
Politics Books  &105 & 441 & 0.559 &0.680 &0.857
& 0.534 &0.645 &0.829
&0.508 &0.638 &0.838
&0.502 &0.516 &0.781
&0.455 &0.594 &0.857\\ \hline
 Amazon &334,863&925,872&0.931 &0.580 &0.998 
&- &- &- 
&- &- &- 
&0.761 &0.092 &0.989
&0.902 &0.490 &0.991\\ \hline
\end{tabular}
\label{tab:real1}
\vspace{-2mm}
\end{table*}

\tabcolsep=4pt
\begin{table*}[tbh]
\centering
\center \caption{Performance of graph clustering algorithms on Large Real-world Networks Without Class information}
\begin{tabular}{|c|c|c|c|c|c|c|c|c|c|c|c|c|c|c|c|c|c|c|c|c|c|c|c|c|c|}
\hline \multirow{2}{*}{Network} &  \multirow{2}{*}{$|V|$} &  \multirow{2}{*}{$|E|$} & \multicolumn{3}{|c|}{Attractor} & \multicolumn{3}{|c|}{Metis} & \multicolumn{3}{|c|}{MCL} \\ \cline{4-12}
  & & & \#C & \emph{mod.} & \emph{cut}  & \#C & \emph{mod.} & \emph{cut} & \#C & \emph{mod.} & \emph{cut} \\ \hline
Collaboration & 9,875 & 25,973
&1384 &0.579 &1179 
&1384 &0.309 &4217 
&2093 &0.537 &2103  \\ \hline 
Friendship & 58228 & 214078
&8045 &0.421 &7325 
&8045 &0.138 &53984 
&13788 &0.319 &36723  \\ \hline 
Amazon & 334,863 & 925,872
&23825 &0.741 &10811 
&23825 &0.451 &47336 
&46557 &0.623 &47488  \\ \hline
Road & 1,088,092 & 1,541,898
&59919 &0.856 &25055 
&59919 &0.673 &31542 
&86745 &0.810 &25065 \\ \hline
\end{tabular}
\label{tab:real2}
\vspace{-2mm}
\end{table*}

\begin{figure}[!tb]
\centering
\begin{tabular}{cccc}
\includegraphics[height=54mm]{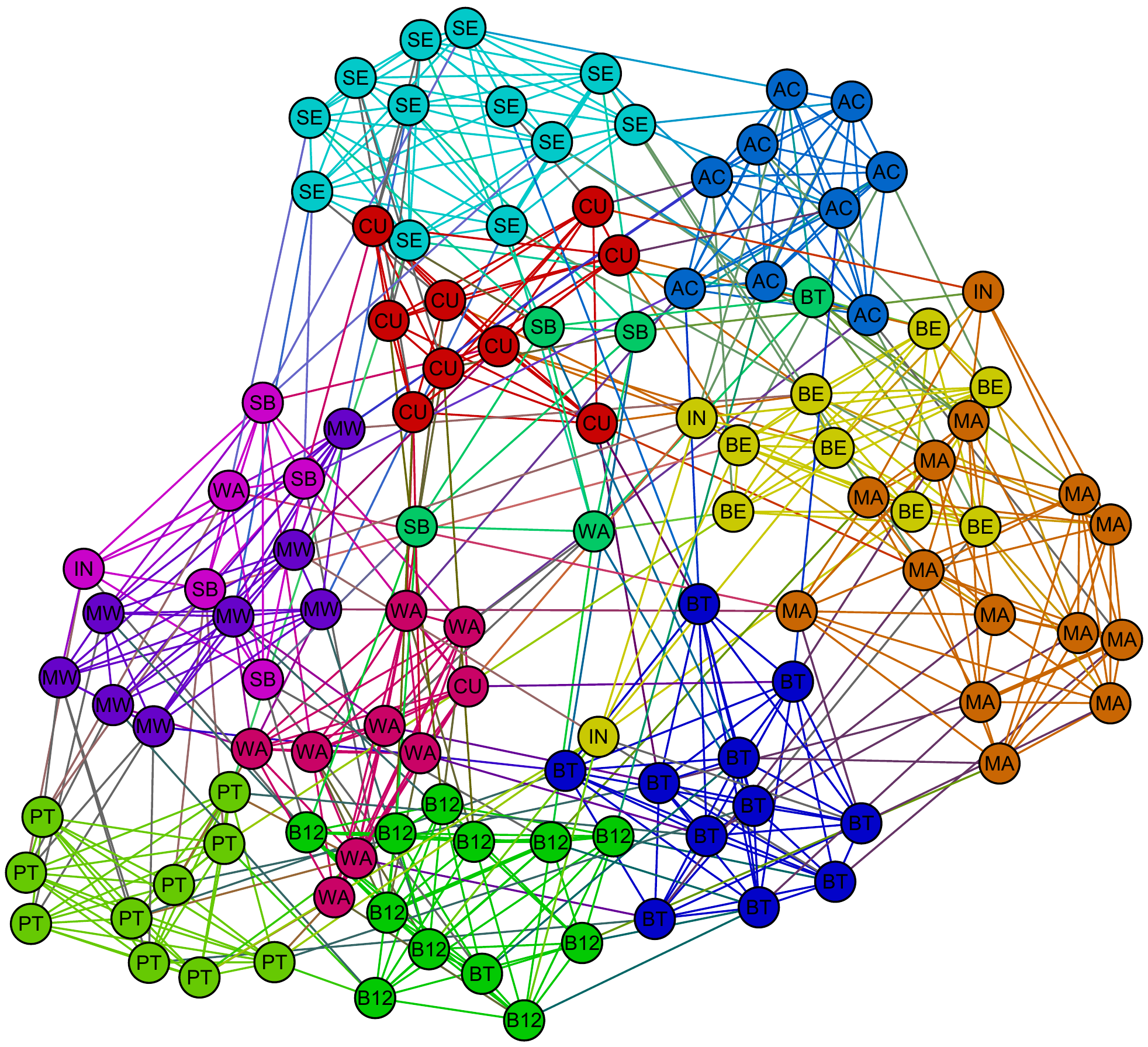}
\end{tabular}
\caption{Performance of Attractor on the American football network.\vspace{-5mm}}
\label{fig:football}
\end{figure}

\subsection{Real World Data}
In this section, we evaluate the performances of different community detection algorithms on small to large real-world networks which are all publicly available from the UCI network data repository (https://networkdata. ics.uci.edu/index.php) and Stanford large network dataset collection (http://snap.stanford.edu/data/).

\vspace{2mm}
\hspace{-6mm}\textbf{(1). Networks with class information}
\vspace{2mm}

We first investigate the networks for which the ground truth of community structure is already known. The external evaluation measures such as \emph{NMI}, \emph{ARI} and \emph{purity} are reported.

 \textbf{Zachary's karate club network:}   The famous network, derived by Zachary's observation about a karate club, reflects the friend relationship among these members.  Specially, the network could be divided into two communities, which reflects the disagreement between the administrator and the instructor.  Fig. \ref{fig:karate} shows that Attractor identifies the communities with a high degree of success (with high values of NMI, ARI and Purity), and outperforms other comparing algorithms (Table \ref{tab:real1}).  Specifically, two communities are successfully found, except one node is viewed as noise (node `10'). It is also interesting to observe that this node is located between two communities, and links with the hub nodes of two communities, respectively. In real-world scenario, it is also difficult to determine its community belonging to. Actually, it is more likely to assign this node to both communities, which is the overlapping clustering that we will not discuss in this study. For comparing algorithms, they also achieve a relatively good performance, and most members are correctly grouped.  The performance of the different algorithms is summarized in Table \ref{tab:real1}.

 \textbf{American college football:} The network derived from the American football games of the schedule of Division I during regular season Fall 2000, where 115 vertices in the graph represent teams, and 613 edges represent regular-season games between the two teams they connect.  The teams are divided into 12 conferences containing around 8-12 teams each, and thereby the real community structure is already known. Fig. \ref{fig:football} plots the communities which are detected by Attractor. It is interesting to note that Attractor automatically finds 12 communities with high quality (NMI = 0.923, ARI = 0.897, Purity = 93.0\%). From this figure, we can observe that most of teams are correctly assigned into corresponding communities. Ncut and MCL find the similar community structure as Attractor. For Metis or Modularity, however, it seems to be more difficult to discover the natural community structure (Table \ref{tab:real1}).

\begin{figure}[tb]
\centering
\begin{tabular}{cccc}
\includegraphics[height=44mm]{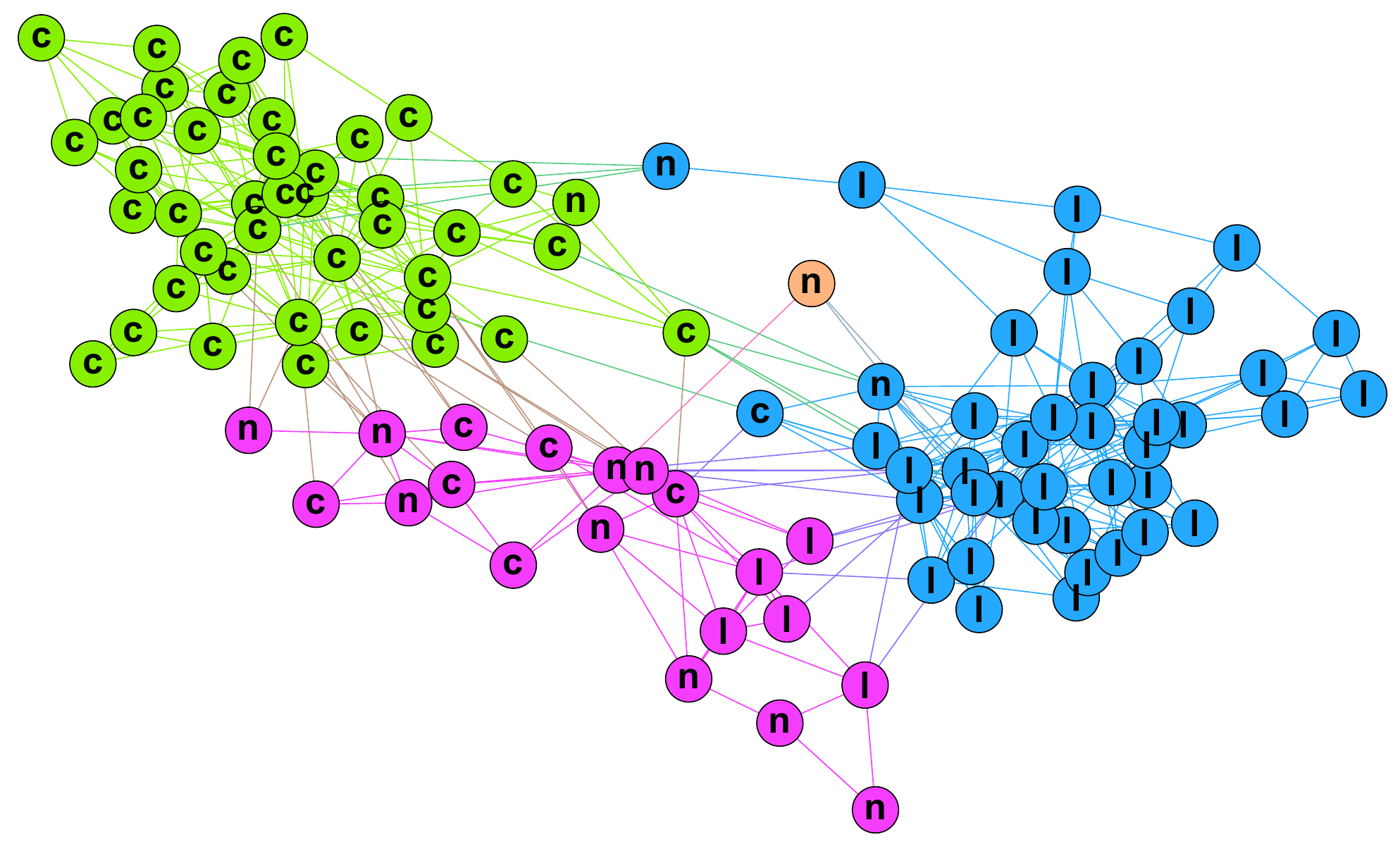}
\end{tabular}
\caption{Performance of Attractor on the network of political books.\vspace{-3mm}}
\label{fig:book}
\end{figure}

 \textbf{Books about US politics:} This network, derived from the politic books about US politics published around the time of the 2004 presidential election, consists of 105 nodes and 441 edges.  Nodes represent books sold by the online bookseller \emph{Amazon.com}.  Edges represent frequent co-purchasing of books by the same buyers.  Each book is labeled with `l', `n', or `c' to indicate whether they are ``liberal", ``neutral", or ``conservative", based on Newman's reading of the descriptions and reviews of the books posted on \emph{Amazon}. Attractor allows a good grouping these books into there categories, where two clusters well represent the corresponding liberal and conservative books, respectively (Fig. \ref{fig:book} and Table \ref{tab:real1}).  For algorithms of Modularity, Metis and Ncut, they produce the comparable results, and most books can be correctly classified. However, MCL yields a relatively bad grouping on this network.

 \textbf{Amazon network:}  This network consists of 334,863 nodes and 925,872 edges, and each node represents a product on the Amazon website.  If two products are frequently co-purchased, there is an edge between them. Each product is categorized to corresponding community based on its category provided by Amazon, and the top 5,000 communities with highest quality were investigated in \cite{groundtruth}. Due to the high time and space complexity of eigenvalue decomposition in Ncut and Modularity, they cannot handle this network, and thus only the results of Attractor, Metis and MCL algorithms on this network are evaluated based on the 5,000 potential ground-truth communities. Attractor obtains the best community quality comparing to other algorithms with high measures (NMI = 0.931, ARI = 0.580, Purity = 0.998) (Table \ref{tab:real1}). Moreover, for comprehensive evaluation, all results of the three algorithms are evaluated by the internal criteria of \emph{modularity} and \emph{cut} (Table \ref{tab:real2}).

\vspace{2mm}
\hspace{-6mm}\textbf{(2). Networks without class information}
\vspace{2mm}

In this section, due to the space and time complexity of Ncut and Modularity, we limit the comparison to the clustering algorithms Metis and MCL on large-scale networks without class ground truth (Table \ref{tab:real2}).

\textbf{Hepth collaboration network:}  The network is a collaboration network of 9,875 authors working on the theory of high energy physics. Attractor identifies 1384 communities, which results in \emph{modularity} = 0.579 and \emph{cut} = 1179.  On the data set, Metis (K = 1384) and MCL also yield a good partitioning while the performance is worse than Attractor in terms of the two measures (Table \ref{tab:real2}).

\textbf{Brightkite friendship network:}  The graph is a location-based friendship network consisting of 58,228 nodes and 214,078 undirected edges.  $Attractor$ finds 8045 communities  and shows a clear advantage over two other algorithms based on the two measures. For Metis (K=8045),  many friends seem to be incorrectly grouped, which result in a low value of \emph{modularity} = 0.138.

\tabcolsep=2pt
\begin{figure}[!tb]
\centering
\begin{tabular}{cccc}
\includegraphics[height=32mm]{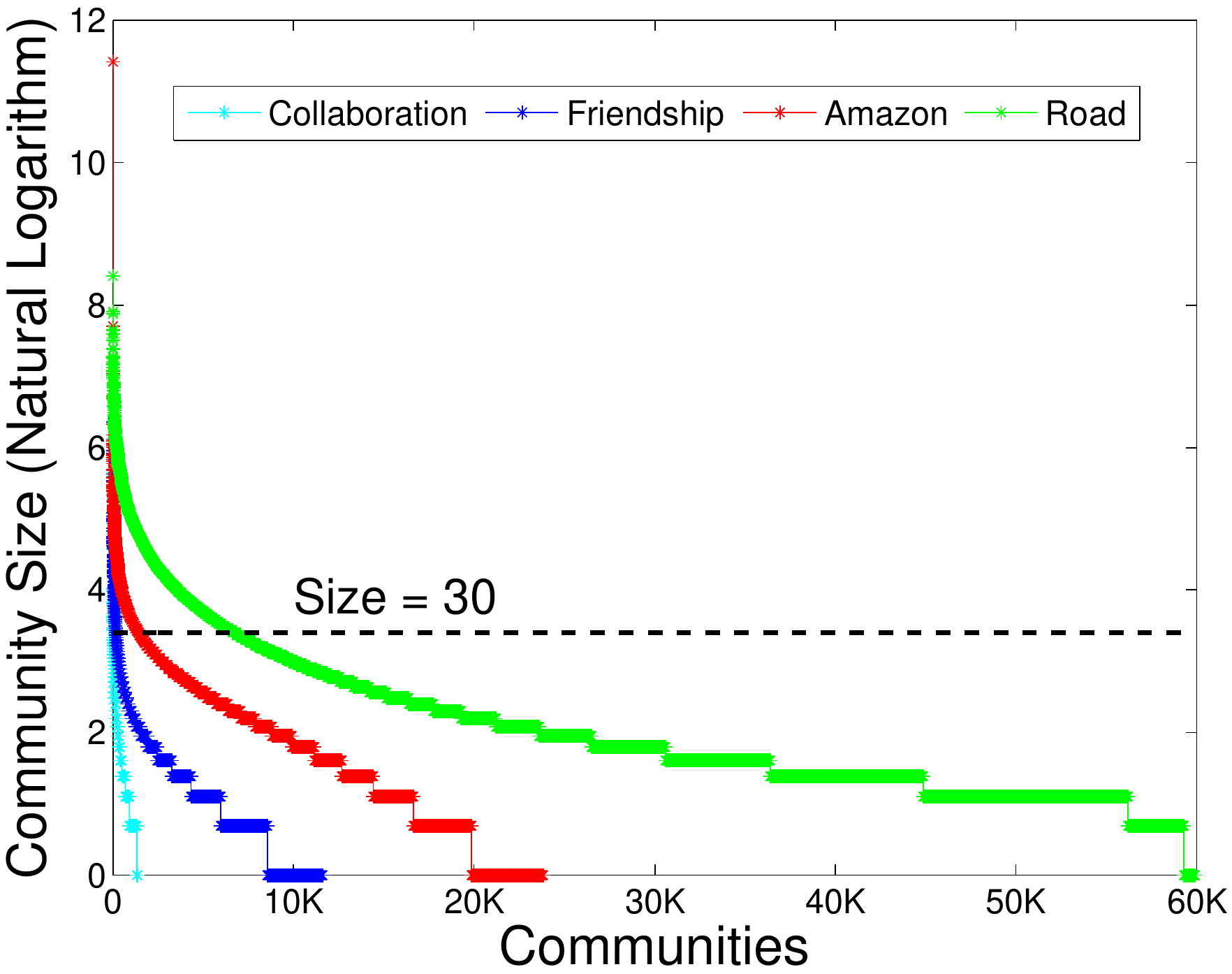}&
\includegraphics[height=32mm]{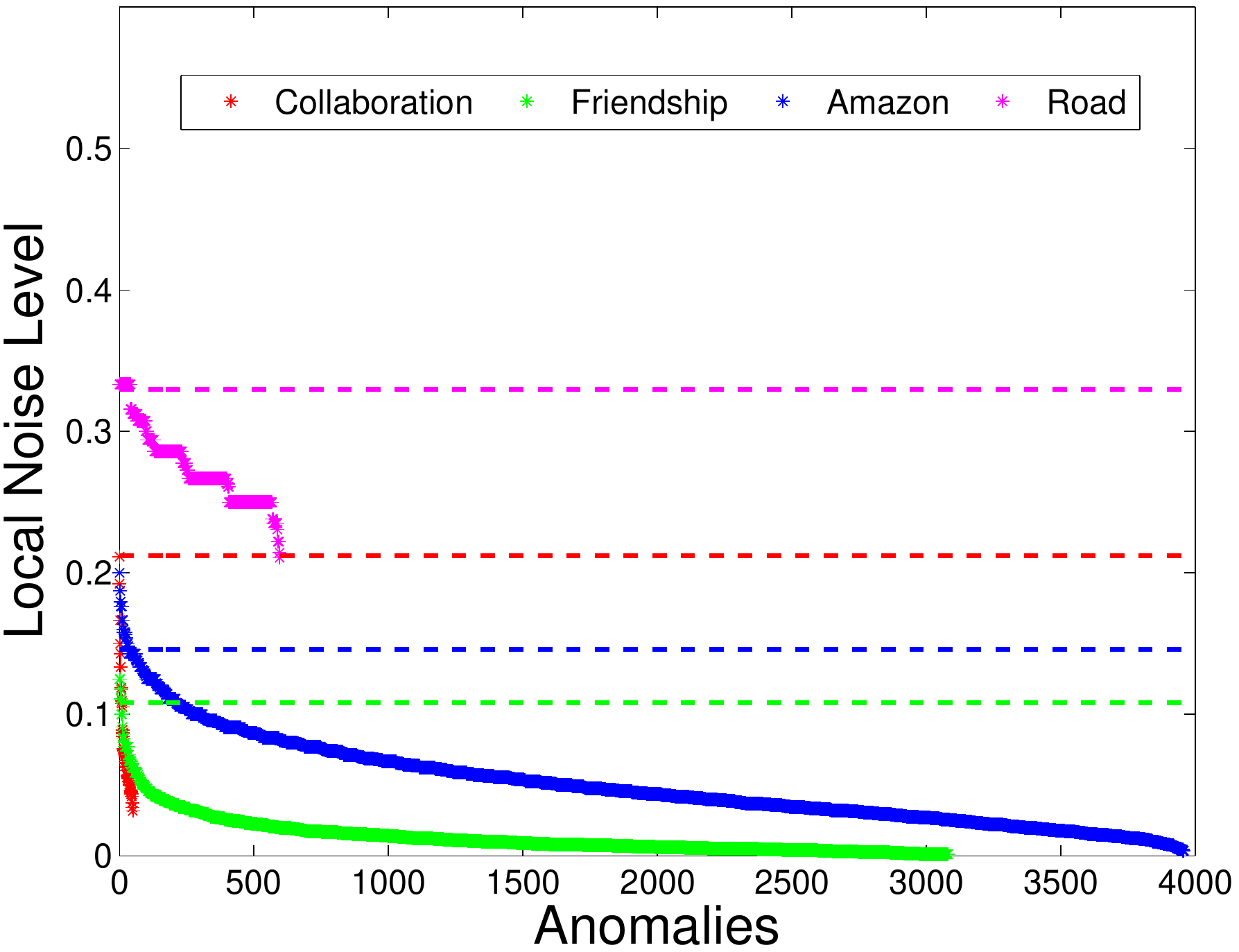}\\
(a) Small communities & (b) Anomalies
\end{tabular}
\caption{Evaluation of small communities and anomalies.\vspace{-3mm}}
\label{fig:distribution}
\end{figure}

\textbf{Pennsylvania road network:}  This network reflects road structure of Pennsylvania, where nodes represent intersections or the endpoints and edges represent the roads connecting these intersections or endpoints.  Here, we set the parameter $\lambda = 0.6$ for Attractor and $i = 1.4$ for MCL as the default values of the two algorithms cannot result in a good results due to the very sparsity of the network.  Attractor finally identifies 59,919 clusters with \emph{modularity} = 0.856 and \emph{cut} = 25055. MCL achieves the comparable performance and is better than the algorithm Metis.

In total, the experiments on the public real-world networks demonstrate that Attractor not only allows extracting meaningful communities in small networks with class label, but also scales up large-scale networks and yields a good graph partitioning in terms of the internal (\emph{modularity} and \emph{cut}) and external measures (NMI, ARI and Purity) (Table \ref{tab:real1}, Table \ref{tab:real2}).

\subsection{Small Community and Anomaly Detection}
\label{sec:noiseHandling}
In this section, we evaluate whether Attractor allows identifying small meaningful communities and anomalies. Fig. \ref{fig:distribution}(a) plots the distribution of community size for the four large real-world networks, and we can see that Attractor can find many small communities. To demonstrate the potential high-quality of the communities, we further examine the quality of the resulting small communities (size$\leq$30) on Amazon network as it has the ground truth for the top 5,000 communities. It is interesting to note that the1458 small communities (size$\leq$30) result in high values of NMI = 0.941, ARI = 0.637 and Purity = 0.989, which potentially shows the desirable property of  small community detection.

Moreover, to check whether the detected anomalies are the potential noisy/unusual nodes, we evaluate the \textbf{local noise level} of each node, which is defined as the fraction of node degree over the number of all links of its neighbors. Fig. \ref{fig:distribution}(b) depicts the local noise level for all resulting anomalies compared with the average noise level (indicated by dashed lines) by Attractor on the four real-world networks,  providing a potential evidence for the effective anomaly detection.

\subsection{Runtime}
\begin{figure}[tb]
\centering
\begin{tabular}{cccc}
\includegraphics[height=54mm]{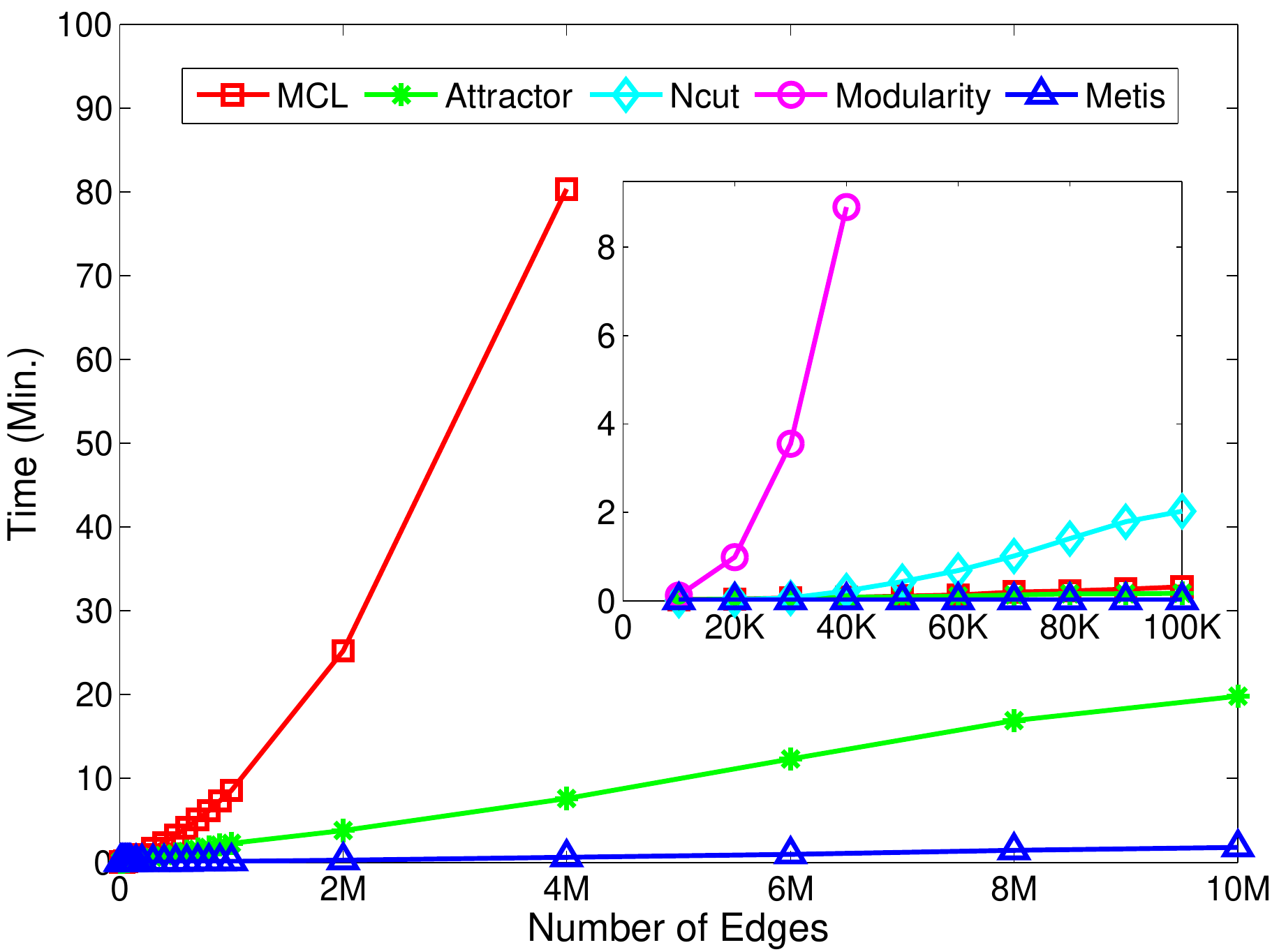}
\end{tabular}
\caption{The runtime of the community detection algorithms. \vspace{-4mm}}
\label{fig:time}
\end{figure}

To assess the scalability of Attractor with respect to network size, we generate several benchmark networks \cite{benchmark} with different edge sizes  ranging from ten thousand to ten million by fixing the average node degree $k = 20$. Fig. \ref{fig:time} shows the running time for different graph clustering algorithms. We can observe that Attractor is faster than Modularity and Ncut since its time complexity is only linear against to $|E|$. Attractor is also faster than MCL, but is slower than Metis.

\tabcolsep=2pt

\section{Conclusions}
\label{sec:conclusion}
In this paper, we introduce Attractor, a new community detection algorithm.  From the distance dynamics point of view, the proposed approach offers an intuitive way to uncover the community structure in networks. Extensive experiments further demonstrate that Attractor allows finding communities on  large  networks with high quality and outperforms several other state-of-the-art community detection methods. In future work, we plan to focus on exploring large network abstraction and visualization based on the intuitive dynamic interaction model.

\end{document}